# Line-based compressive sensing for low-power visual applications

By Mansoor Ebrahim, Sunway University, Malaysia, and Syed Hasan Adil, Daniyal Nawaz and Kamran Raza, Iqra University, Pakistan

In digital signal processing systems, images are usually first transformed into digital signals and then compressed using standard codec algorithms like JPEG, JPEG 2000, MPEG and others. Most of these conventional algorithms require a significant amount of processing and hence computing power, which increases the encoder's energy consumption, making them unsuitable for low-power applications such as wireless or visual sensor networks.

Recently a new method called compressive sensing (CS) has been shown to be more efficient at complex processing at low power. With CS, computation is shifted from the encoder to the decoder – a direct opposite of conventional approaches.

Conventional methodologies such as JPEG, JPEG 2000, MPEG and H.264 use the Shannon sampling theorem, or the so-called Nyquist rate, to transform signals, whereas with CS a signal is represented by a few non-zero coefficients – fewer than the Nyquist rate. The CS scheme effectively decreases the computational requirements (memory, processing power and transmission bandwidth) at the encoder, by combining into a single process the signal acquisition (sampling) and dimensionality (the amount of data that will stream out).

### A New Method

In this article we propose a line-based sampling approach for visual applications using CS, for fast, efficient and less computationally-complex sampling of images.

With our method, the original image is first divided into N multiple lines of the same size, with each line processed independently using the sampling operator Φ. Such an approach benefits CS because:

(i) line-based measurement is faster for practical applications, since sampled image data need not be encoded in its entirety but line by line, until sampling of the whole image is complete;

(ii) practical implementation and storage of the sampling operator are simpler because they deal with a minimum number of samples;

(iii) the individual processing of each image-data block results in an easy solution with a significantly faster and better

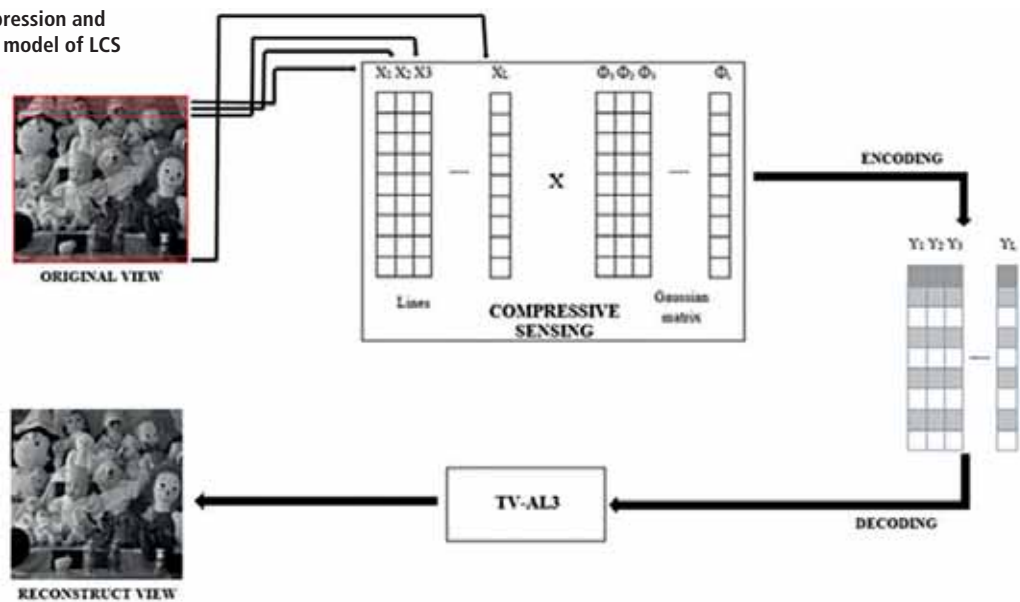

**Figure 1:** Compression and reconstruction model of LCS

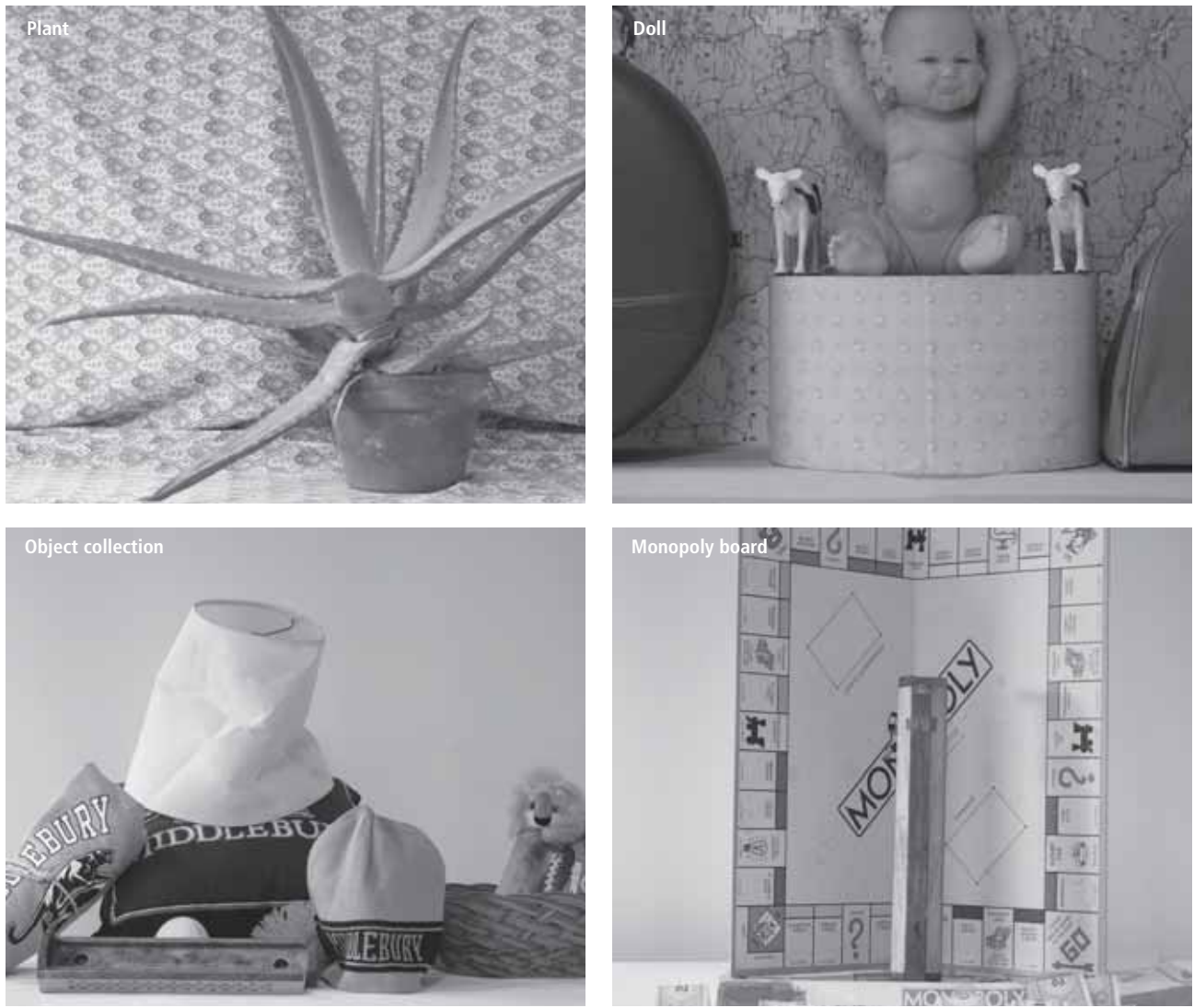

**Figure 2: Several standard grayscale test images of 512 x 512 size**

reconstruction process.

The CS method relies on two important parameters:
- Sparsity, which is important for sampling and reconstruction of a signal; and
- Incoherence (sensing modality), which helps determine the maximum correlation measured between any two elements from two different matrices.

If we consider a signal X with length N to be recovered from M measurements (M << N), sparse in some transformation domain $\Psi$, with random measurement matrix $\Phi$, the set of measurements y is then:

$$y = \Phi X \quad (1)$$

where $X \in R_N$ is the input signal and $y \in R_M$ is the measurement vector. It is assumed that the random sampling matrix $\Phi$ is orthonormal (a matrix with a transpose equal to its inverse, or 1), i.e. $\Phi \times \Phi^T = A$, where A is the identity matrix, or unit matrix.

As the number of unknowns is much larger than the number of observations, recovery of $X \in R_N$ from its corresponding $y \in R_M$, i.e. inverse projection of $X' = \Phi^{-1} y$, is not sufficient.

In our approach, the line-based encoded image is reconstructed by using Total Variation (TV) minimisation, which uses piece-wise smooth characteristics of the signals rather than finding the sparse solution in the transformation domain. The basic TV minimisation function is given as:

$$TV(X) = \sum i,j \; ||Xi+1,j - Xi,j|| + ||Xi,j+1 - Xi,j|| \quad (2)$$

$$min_x \; X \; E||y - \Theta X|| + \lambda \; TV(X) \text{ subject to } \Theta = \Phi \Psi \quad (3)$$

where E is the $\ell_2$ norm (also known as 'least squares', used to minimise the sum of squares of differences between the target and estimated values). However, the basic TV minimisation CS reconstruction problem in Equation 3 is exposed to additional computational burden, i.e. memory usage, processing and transmission power, restricting its use for CS reconstruction. There's a scheme called TV-AL3 that can solve this equation, which combines the conventional Augmented Lagrangian (AL) method with variable-splitting and alternating-direction methods. The TV-AL3 depends on global structurally-random matrices (mainly used for producing fast and efficient sensing matrices in CS measurements), and can generate the same high-quality reconstructed image as the standard Total Variation method but with less processing.

| Sampling Rate | 0.05 | 0.1 | 0.15 | 0.2 | 0.25 | 0.3 |
|---|---|---|---|---|---|---|
| **Plant** | | | | | | |
| **Conventional CS scheme** | 23.93 | 24.95 | 26.32 | 27.52 | 28.45 | 29.32 |
| **Proposed scheme** | 24.76 | 26.16 | 27.54 | 28.78 | 29.87 | 30.86 |
| **Doll** | | | | | | |
| **Conventional CS scheme** | 24.29 | 29.31 | 31.83 | 33.47 | 35.2 | 36.6 |
| **Proposed scheme** | 25.75 | 31.38 | 34.18 | 36.09 | 37.92 | 39.39 |
| **Monopoly board** | | | | | | |
| **Conventional CS scheme** | 23.7 | 27.4 | 29.67 | 32.65 | 35.07 | 37.34 |
| **Proposed scheme** | 24.54 | 28.38 | 30.88 | 34.01 | 36.47 | 38.93 |
| **Object collection** | | | | | | |
| **Conventional CS scheme** | 24.53 | 27.53 | 29.44 | 32.05 | 33.98 | 35.7 |
| **Proposed scheme** | 26.74 | 30.19 | 32.71 | 35.42 | 37.49 | 39.33 |

**Table 1: R-D performance (dB) achieved by five trials of a previous scholarly scheme and our proposed scheme for different images**

### Line-Based Compressive Sensing

Consider an $I_R \times I_C$ image captured by a visal node, where $I_R$ and $I_C$ are the total number of pixels in each row and column, respectively. At the encoder, the proposed line-based CS is applied to the image, which first has been divided into N multiple lines (each 1 x L in size), and then each line processed independently using the sampling operator $\Phi$.

If $X_i$ is the vectorised signal of the i$^{th}$ line of the image, then the compressed CS vector output $Y_i$ is:

$$min_X ||D_m X||_n, \text{ subject to (s.t.) } Y_i = \Phi_L X_i \quad (4)$$

where $D \in (D_X, D_Y)$ are the horizontal and vertical gradients respectively, and $\Phi_L$ is an ortho-normalised, independent, identically-distributed Gaussian matrix.

For the whole image, each line is then individually sampled using the same measurement matrix $\Phi_L$ with a constrained structure, or the optimal solution. The measurement Y is then transmitted to the decoder for reconstruction. The encoding process then goes as follows:

**LCS ENCODER**

**Input: Grey scale image (2-dimensional image);**

**Image to lines (%)**
**Consider an image I**
**for r = 1 to $r_{Max}$**
**for c = 1 to $c_{Max}$**
**re-arranges each distinct block $I_{rc}$ into a column of X.**

**Compressive sensing (%)**
**Now consider vector X (lines)**
**for i = 1 to $i_{Max}$**
**for each line i, sampled with measurement matrix $\Phi$**
**$Y_i = \Phi X_i$**
**Output: Y, the encoded sample**

At the decoder, the line-based encoded measurements Yi of the image are decoded by solving the TV minimisation problem of Equation 3, using the AL method with variable splitting, and alternating it with the direction method, to reconstruct the original image:

$$TV-AL3(X) = min_X ||W_m||_n + \lambda ||X-AX||^2 \quad (5)$$

subject to $W_m = D_m X$, A = Augmented Lagrangian filter
where $\lambda$ = extra plenty parameter.

The proposed decoding process of the encoded sample is defined as follows:

**DECODER**

**Input:**
**Y = encoded sample;**
**$\Phi$ = measurement Gaussian matrix;**

Figure 3: Visual quality comparison for reconstruction of the monopoly board image at sampling rates of 0.1 using (a) conventional CS method; and (b) the proposed line-based CS method

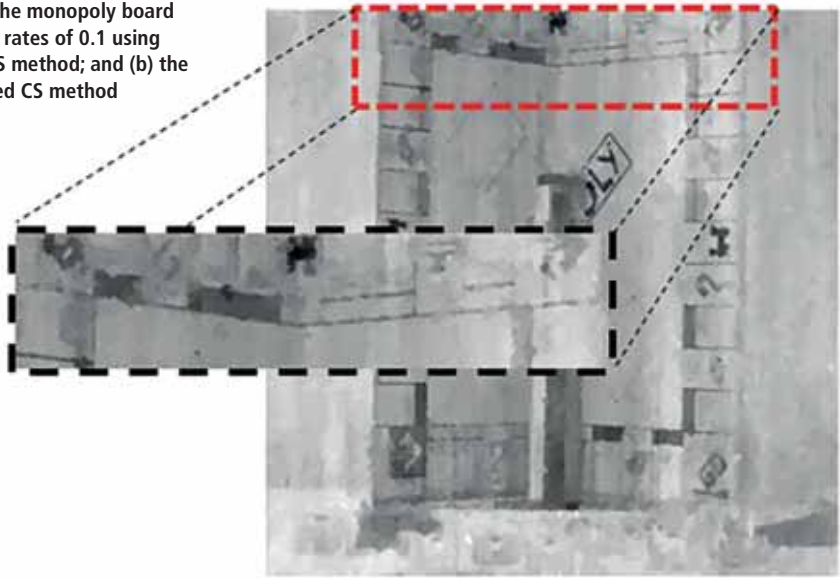

a) Visual results of conventional compressive sensing

R = number of rows of the actual image;
C = number of columns of the actual image,
λ = Plenty Parameter (line size, compressive ratio)

**Rearrangement (%)**
for r = 1 to R
for c = 1 to C
re-arranges each block of encoded image into a column of X.

**Reconstruction of the encoded image of size R x C (%)**
Now consider vector X
        I <- TVAL3(X, λ)
Output: I, the reconstruction of the encoded sample image

### Improved Images

The overall process of line-based compression and reconstruction is shown in Figure 1.

The proposed scheme is implemented using Matlab version 8.3.0.532 (R2014b) on an Intel Xeon CPU E5-1620 desktop computer with a 3.6GHz processor and 8GB RAM. Its performance was evaluated with a set of standard greyscale images 512 x 512 in size; see Figure 2.

The evaluation was carried out by measuring the rate distortion (R-D) in terms of peak signal-to-noise ratio, or PSNR (dB), for different sampling rates. Because of the random nature of the measurement matrix Φ, the quality of the reconstructed image varies, so we did five independent trials. All images are encoded at sampling rates of 0.05, 0.1, 0.15, 0.2, 0.25 and 0.3; see Table 1.

Table 1 shows that the proposed scheme is better than the conventional CS scheme, with improved gain of 1-3dB, for example. Moreover, for more complex images such as 'object collections' the performance gain at 2-3dB is even higher.

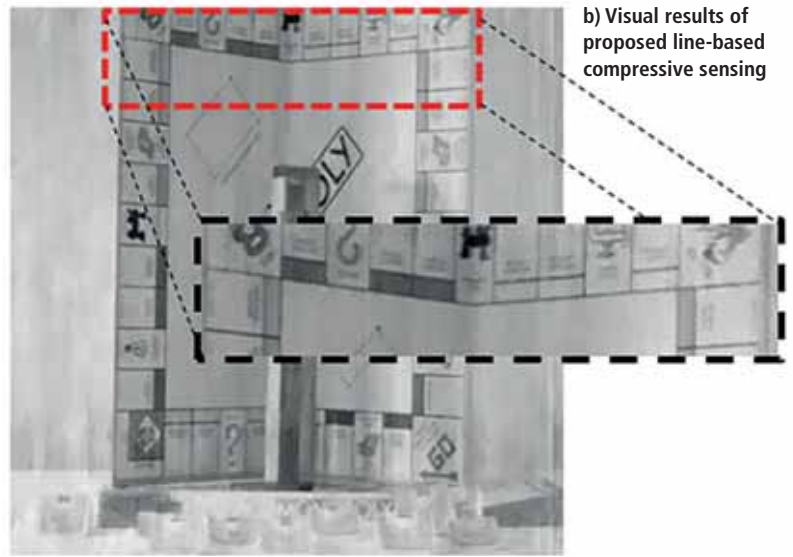

b) Visual results of proposed line-based compressive sensing

By comparing the visual results in Figure 3, it can be seen that the image reconstructed using the proposed scheme improves the blurring in the image reconstructed by the conventional CS method. Equally, by comparing the highlighted regions (red dotted area), it can be seen that the image reconstructed using our scheme is much sharper than the conventional one.

### Fast and Efficient

This study presents a line-based compressed sensing scheme for low-power visual applications. The scheme is simpler, provides faster and more efficient initial recovery solution of images, than other, conventional methods.

However, there is further work to be done, especially on the optimisation criteria of the line-based sampling operator, and the comparison between this method with the block-based CS scheme. ●